\documentclass[useAMS,usenatbib]{mn2e}

\usepackage{times}
\usepackage{graphics,epsfig}
\usepackage{graphicx}
\usepackage{amsmath}
\usepackage{amssymb}

\newcommand{\arcs}{$^{\prime\prime}$}
\newcommand{\arcm}{$^{\prime}$}
\newcommand{\ha}{\ensuremath{{\rm H}\alpha}}
\newcommand{\ms}{\ensuremath{\rm M_{\odot}}}

\newcommand{\um}{\ensuremath{\rm \mu m}}

\title[Radio-FIR correlation in dwarfs]{The radio - far infrared correlation in the faintest star forming dwarf galaxies}
\author[Roychowdhury \& Chengalur]{
Sambit Roychowdhury$^{1}$\thanks{E-mail: sambit@ncra.tifr.res.in (SR); chengalu@ncra.tifr.res.in (JNC)} and Jayaram N. Chengalur$^{1\star}$\\
       \\
       $^{1}$NCRA-TIFR, Post Bag 3, Ganeshkhind, Pune 411 007, India}

\begin{document}
\date{}

\pagerange{\pageref{firstpage}--\pageref{lastpage}} \pubyear{}

\maketitle

\label{firstpage}

\begin{abstract}
We study the radio - far-infrared (FIR) correlation in a sample of faint dwarf irregular galaxies using NVSS data for 1.4 GHz radio flux, {\emph Spitzer} MIPS 70\um~data for FIR flux, and {\emph GALEX} FUV data to estimate the star formation rates (SFR). Since our target galaxies are extremely faint, we  stack images of many galaxies together to estimate the average radio and FIR fluxes.  We find that for a given SFR both 70\um~and 1.4 GHz fluxes are low compared to the calibration for large spirals. Nonetheless, the ratio of 70\um~to 1.4 GHz flux agrees within errorbars with that seen for large galaxies. The radio-FIR correlation thus appears to be the result of a `conspiracy'. We use the SFR to estimate the non-thermal fraction of the 1.4 GHz radio emission and find it to be around 50\%, much smaller than the 90\% typical for spirals. We also estimate the equipartition magnetic field and find it to be $\sim$ 2 $\mu$G, about five times smaller than that typical for spirals.
\end{abstract}


\section{Introduction}
\label{sec:int}

Despite many decades of study, \citep[see eg.][]{con92}, the origin of tight correlation between global radio and far infrared (FIR) flux of normal star-forming galaxies  is still not well understood.  The correlation has been observed to exist over five orders of magnitude in both radio and FIR luminosity \citep{yun01} and exists even at intermediate redshifts \citep{app04}. The correlation has been suggested to be a consequence of both the 1.4~GHz flux and the FIR flux being separately correlated to the star formation rate (SFR). The radio 1.4~GHz emission comes mainly from synchrotron emission from cosmic ray (CR) electrons accelerated in supernova remnants (SNR) interacting with the magnetic field, and the FIR flux comes mainly from dust re-processing of UV photons from young stars; both are hence correlated to the SFR. However, while it is true that both the radio and the FIR fluxes are expected to correlate to the SFR and hence to each other, the tightness of correlation for disparate systems with large differences in magnetic field strengths remains difficult to understand.

Several models have been proposed to explain this tight correlation.  In the early `calorimeter' model \citep{vol89}, galaxies are assumed to be optically thick to both UV photons and CR electrons, and the energies from both are re-processed completely within the galaxies themselves. The model also assumes source strengths of CR electrons and ionizing photons are both proportional to the SN rate, and that the energy densities of the interstellar radiation field and the magnetic field in all galaxies is a constant. While this model produces a tight radio-FIR correlation, it also predicts radio spectra from star-forming galaxies that are steeper than observed \citep[e.g.] []{con92}. \citet{tho06} resolves this anomaly by invoking ionization and bremsstrahlung losses at low radio frequencies, and also suggests that magnetic fields in galaxies are much higher than the equipartition field, contributing to the CR electrons loosing their energy in a short timescale. An alternative `optically thin ISM' explanation is proposed by \citet{hel93}, in which most of the UV photons and relativistic CR electrons escape the galaxy disc, the latter because of a proposed correlation between disc scaleheight and escape scalelength of CR electrons. \citet{mur05} present another model in which the radio continuum, CO and FIR emissions are all determined by the hydrostatic pressure in galaxies instead of the SFR. However, despite the large amount of modeling effort, a consensus on the cause of the tight radio-FIR correlation remains elusive.

Most of the above mentioned studies focus on large star forming galaxies. A specific model which includes small dwarf galaxies was presented by \citet{bel03}. In this model a `conspiracy' maintains the radio-FIR correlation for low luminosity galaxies, viz. the FIR emission reduces because of the low dust content of these galaxies, while the non-thermal radio emission becomes inefficient due to the easy escape of CR electrons. A more recent numerical work by \citet{lac10} attributes the radio-FIR correlation to `calorimetry', combined with two different conspiracies at high and low gas density regimes. At low gas densities more CR electrons escape while low UV opacity causes lower FIR emission.

Measuring the radio continuum flux from the faintest dwarf irregular galaxies is interesting even apart from the issue of the radio-FIR correlation. \citet{pad11} predict that star-forming dwarfs will contribute significantly to the number counts at faint levels in the proposed next generation deep sub-$\mu$Jy surveys. Measuring the radio continuum flux also allows one to estimate the magnetic fields strengths.  Magnetic fields in low mass star-forming dwarfs can give clues to understanding the seeding of the intergalactic medium (IGM) at early epochs by similar low mass galaxies, due to easy escape of material from their shallow gravitational potentials \citep{ber06}. Magnetic field strengths comparable to those in normal spiral galaxies have been observed for nearby dwarf galaxies with extreme properties (i.e. evolved objects with recent or ongoing starbursts) like NGC 4459, NGC 1569, NGC 6822, IC 10 \citep{chy00,kep10,chy03,chy11}.
However for faint dwarf irregular galaxies, the available data is very scarce. The SMC has been measured to have a large-scale weak magnetic field of $\sim$3 $\mu$G \citep{mao08}. A systematic survey of radio continuum emission from local group dwarf irregular galaxies \citep{chy11} resulted in only 3 new detections with magnetic fields estimated to be $<$~5 $\mu$G.

The dearth of systematic studies of the radio-FIR correlation in faint dwarf irregular galaxies is in large part because of the difficulty of detecting radio continuum emission from them. We present here a study of the radio-FIR correlation where the radio and FIR fluxes are obtained from stacked images of the individual galaxies. Our sample galaxies are chosen from the Faint Irregular Galaxy GMRT Survey (FIGGS) \citep{beg08} sample, which is a systematically selected sample of faint, gas-rich, star-forming dwarf galaxies. Details of the radio, FIR and FUV fluxes are presented in section~\ref{sec:pro}. In Sec.~\ref{sec:dis} we discuss the implication of our results for the radio-FIR correlation and the magnetic field strength. A summary of our findings is given is Sec.~\ref{sec:sum}.

\section{Procedure and Results}
\label{sec:pro}

As described above, we have available to us 70$\mu$m and 1.4~GHz fluxes for our sample of faint dwarfs. In order to check the linearity of the radio-FIR correlation, we would need a control sample of large galaxies for which the radio-FIR correlation has been measured using fluxes in these same bands. A suitable sample is provided by the sample of \citet{app04}. These authors give the radio-FIR correlation for a large sample of galaxies in terms of q70, defined as log${\rm _{10}({S_{70\mu m}}/{S_{1.4 GHz}})}$, where S$_{\rm 70\mu m}$~and S$_{\rm 1.4 GHz}$~represent the fluxes as measured by the {\emph Spitzer} MIPS 70\um~band and at 1.4 GHz respectively. We compute below the value of q70 for our own sample and compare it with that found by \citet{app04}.

As mentioned above, our sample of dwarf galaxies is drawn from the
FIGGS survey. Although GMRT radio continuum data is available for the 
galaxies (from the line free region of the spectrum) the bandwidths
are narrow and the sensitivity is low. The radio continuum data
was hence instead taken from the NRAO VLA Sky Survey (NVSS) \citep{con98}. The NVSS has an effective continuum bandwidth of $\sim$42 MHz, synthesised beam FWHM (angular resolution) of $\sim$45\arcs, and background rms (sensitivity) of $\sim$0.45 mJy bm$^{\rm -1}$. Far infrared (FIR) fluxes for the galaxies were obtained from archival {\emph Spitzer} \footnote{This work is based [in part] on observations made with the Spitzer Space Telescope, obtained from the NASA/ IPAC Infrared Science Archive, both of which are operated by the Jet Propulsion Laboratory, California Institute of Technology under a contract with the National Aeronautics and Space Administration.} Multiband Imaging Photometer Spitzer (MIPS) \citep{rie04} 70\um~images. The 70\um~band of MIPS has a bandwidth of $\sim$19\um, and an angular resolution of $\sim$19\arcs. The star formation rate (SFR) was estimated from 
archival \emph{GALEX} \footnote{Some of the data presented in this report were obtained from the Multimission Archive at the Space Telescope Science Institute (MAST). STScI is operated by the Association of Universities for Research in Astronomy, Inc., under NASA contract NAS5-26555. Support for MAST for non-HST data is provided by the NASA Office of Space Science via grant NAG5-7584 and by other grants and contracts.} FUV band data. The FUV data have a resolution of
$\sim 4$\arcs.

Although all FIGGS sample galaxies have been covered by the NVSS, not all of the NVSS data are useful for the current purpose. Galaxies where there is a background continuum source superposed on the image cannot be used, since their inclusion would bias the radio flux upwards. Visual inspection of all of the NVSS images left us with 57 galaxies that have no strong background source near the galaxy.
MIPS 70\um~data is available for 26 of the FIGGS galaxies, 13 of these galaxies have detectable emission in the 70\um~band. For all of the detected galaxies the  70\um~flux is less that 2\arcm~in extent.
Therefore we decided to use the standard MIPS pipeline (filtered and mosaiced) processed basic calibrated datasets (PBCDs). Multiple PBCDs for the same galaxy, wherever available, were processed separately.
Public \emph{GALEX} FUV data is available for 46 of the FIGGS galaxies. 

As described above, data at different bands is available for different subsets of the FIGGS galaxies. We hence define the following subsamples whose properties we will examine in the following analysis.
\begin{enumerate}
\item `NVSS' sub-sample consists of the 57 galaxies with     available radio continuum data.
\item `MIPS 70\um' sub-sample consists of the 26 galaxies with available FIR data.
\item `FUV' sub-sample consists of the 46 galaxies for which {\emph GALEX} FUV data is available.
\item `Common' sub-sample consists of 24 galaxies for which 1.4 GHz, MIPS 70\um, and FUV fluxes are available.
\end{enumerate}
Only one galaxy, UGC 5456 is detected at 1.4 GHz in NVSS, and also has detectable 70\um~ and {\emph GALEX} FUV emission. The properties of the three subsamples as well as the galaxy UGC 5456, are given in Table~\ref{tab:samp}. As can be seen the mean properties
of all of the sub-samples are quite similar, and not surprisingly our
results are not very sensitive to which exact sub-sample (or which mixture of sub-samples) is selected for the analysis. 

\begin{table}
\begin{center}
\caption{Sample Properties}
\label{tab:samp}
\begin{tabular}{|lccccc|}
\hline
Sample/Galaxy&number  of&M${\rm{_{B}}}^1$&D${_{\rm Ho}}^2$&M${\rm{_{HI}}}^3$&D$^4$\\
             & galaxies &                &  ($^\prime$)   &    (10$^7$\ms)  &(Mpc)\\
\hline
\hline
NVSS&57&$-$13.1&1.7&2.8&4.8\\
MIPS 70\um&26&$-$13.1&2.0&2.6&3.4\\
FUV&46&$-$13.1&1.7&2.6&4.5\\
common&24&$-$13.1&2.0&2.2&3.4\\
UGC 5456&1&$-$15.1&1.9&5.9&5.6\\
\hline
\hline
\end{tabular}
\end{center}
\begin{flushleft}
$^1$ Absolute blue magnitude, median value for the subsamples; $^2$ Holmberg diameter, mean value for the subsamples; $^3$ total HI mass, median value for the subsamples; $^4$ distance, mean value for the subsamples.
\end{flushleft}
\end{table}

Preliminary processing of the images were done using the {\it Astronomical Image Processing System} (AIPS).
Available MIPS 70\um~PBCDs have pixel sizes of 4\arcs, and hence NVSS cutout images were obtained with the same pixel sizes.
The NVSS cutout images and the MIPS 70\um~PBCDs for each galaxy were aligned and the image sizes were made identical using the AIPS task HGEOM, with the galaxy at the centre of each field.
The continuum sources visible (other than the galaxy itself, if detected) were identified visually and blanked out using the AIPS task BLANK. The problem of flux lying in the extended wings of the MIPS 70\um~beam was removed by convolving each blanked PBCD with the suitable kernel from \citet{ani11}, to give a convolved image with a Gaussian beam of FWHM 41\arcs (comparable to the NVSS beam FWHM).
The convolution was done using the AIPS task CONVL, and the extracted flux shows about a 10\% increase after such a convolution is done. Calibration uncertainties for the MIPS 70\um~band is about 5\% \citep{gor07}, and this 5\% was taken to be the floor of the measurement errors. Using the values from \citet{sch98}, assuming ${\rm A_V/E(B-V)}$~= 3.1, and the reddening curve from \citet{li01}, we find that the change in flux due to Galactic extinction even in the direction of maximum Galactic extinction (towards the galaxy KKH 98) is less than 1\%. The uncertainties in the measured FIR due to other factors is much more, therefore we ignore the correction due to Galactic extinction.

\begin{figure}
\psfig{file=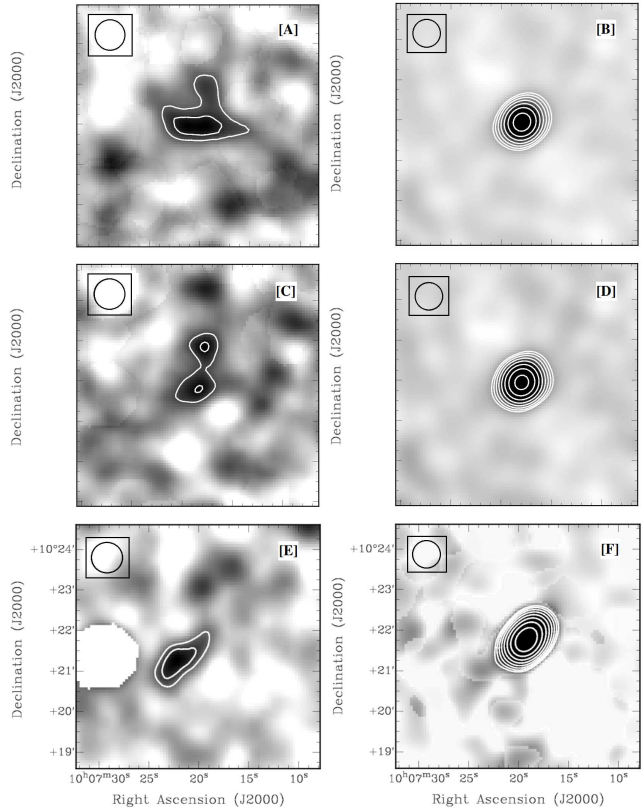,width=3.5truein}
\caption{[A] and [B] are the stacked 1.4 GHz and 70\um~images from the `NVSS' and `MIPS 70\um'~sub-samples respectively. [C] and [D] are the stacked 1.4 GHz and 70\um~images from the `common' sub-sample. [E] and [F] are the 1.4 GHz and 70\um~images of UGC 5456. Each image is a $\sim$6\arcm$\times$6\arcm~field, with the corresponding beam sizes shown. For images [A], [C] and [E], the contours are at 2$\sigma$ (only the contour around the source) and 3$\sigma$ levels. For images [B] and [D], the first contour is at 3$\sigma$ level, and for image [F] it is at 5$\sigma$ level. For images [B], [D] and [F], consecutive contours are in intervals of $\sqrt{\rm 2}$.}
\label{fig:im}
\end{figure}

The radio maps for the `NVSS' and `common' sub-samples, as well as the FIR PBCDs for the `MIPS 70\um' and `common' sub-samples, were stacked separately. The images were co-added after being weighted by the inverse of the variance of the flux in the background pixels.
The co-added images for various subsamples, as well as those for the galaxy UGC 5456, are shown in Figure~\ref{fig:im}.
The background rms level of the stacked `NVSS' sub-sample radio image, obtained after co-adding 57 images, was $\sim$66 $\mu$Jy bm$^{\rm -1}$. Correspondingly, the background rms level of the stacked `MIPS 70\um' sub-sample image, obtained after co-adding 51 PBCDs with median background rms level $\sim$0.2 MJy sr$^{\rm -1}$, was $\sim$0.03 MJy sr$^{\rm -1}$. As expected, stacking N images together resulted in the background rms level going down as $\sqrt{\rm N}$.

Fluxes were extracted from the 1.4 GHz and 70\um~images using the following procedure. Isophotes were fitted to the image using the IRAF image processing software, and the isophote where the mean flux value falls to the background flux level is identified. The total flux in each case was extracted by summing over the flux in all the pixels included within the above mentioned isophote, using the AIPS task BLSUM. The fluxes thus extracted are listed in Table~\ref{tab:flux}.
Note that that the fluxes of the stacked radio sources obtained from the `NVSS' and `common' sub-samples, and the fluxes of the stacked FIR sources obtained from the `MIPS 70\um' and `common' sub-samples, agree within the measurement uncertainties.

\begin{table}
\begin{center}
\caption{calculated fluxes and q70}
\label{tab:flux}
\begin{tabular}{|lccc|}
\hline
Sample/galaxy&1.4 GHz flux&70\um~flux&q70\\
             &(mJy)&(mJy)&\\
\hline
\hline
NVSS&0.9$\pm$0.2&&2.0$\pm$0.2$^*$\\
MIPS 70\um&&83$\pm$5&\\
common&0.8$\pm$0.3&90$\pm$8&2.0$\pm$0.4\\
UGC5456&3$\pm$1&560$\pm$30&2.3$\pm$0.3\\
\hline
\hline
\end{tabular}
\end{center}
\begin{flushleft}
$^*$ Using FIR flux from `MIPS 70\um' sub-sample and radio flux from `NVSS' sub-sample
\end{flushleft}
\end{table}

For the `common' sub-sample, the q70 value is $2.0 \pm 0.4$. If one assumes that the `NVSS' sub-sample gives a fair measure of the average 1.4 GHz flux of the FIGGS galaxies, and the `MIPS 70\um' sub-sample gives a fair measure of the 70\um~ flux of the FIGGS galaxies, then the q70 for the FIGGS sample as a whole is $\sim 2.0 \pm 0.2$. For UGC 5456 alone the q70 is $2.3 \pm 0.3$. All of these values (which are also listed in Table~\ref{tab:flux}) agree within the error bars. Interestingly, they also agree with the value of $2.15 \pm 0.16$ obtained by \citet{app04} for a sample of large galaxies. The value
quoted by \citet{app04} is the mean and dispersion over the q70 values
for the individual galaxies in his sample. On the other hand, the q70
values we quote (leaving aside the q70 estimate for UGC 5456) are for the ratio of the means of the 1.4 GHz and 70$\mu$ fluxes for the galaxies in various sub-samples. We hence computed the same quantity using the data from \citet{app04}. The value we get $1.99 \pm 0.17$~(where the error bars have been computed using bootstrap re-sampling) agrees within the error bars with the values we get for the different FIGGS sub-samples.

\section{Discussion}
\label{sec:dis}

\begin{table*}
\begin{center}
\caption{calculations based on deduced fluxes}
\label{tab:cal}
\begin{tabular}{|lcccccccccc}
\hline
Sample/$^a$&SFR$_{\rm FUV}^b$&L$_{\rm H\alpha}^c$&L$_{\rm 70~\mu m}^d$&L$_{\rm 70~\mu m}^{{\rm high Z}, e}$&L$_{\rm 1.4~GHz}^f$&L$_{\rm 1.4~GHz}^{{\rm >L_*}, g}$&L$_{\rm 1.4~GHz}^{{\rm <L_*}, h}$&L$_{\rm thermal}^i$&non-thermal$^j$&B$^k$\\
galaxy&(M$_{\odot}$~yr$^{\rm -1}$)&(ergs s$^{\rm -1}$)&(ergs s$^{\rm -1}$)&(ergs s$^{\rm -1}$)&(W Hz$^{\rm -1}$)&(W Hz$^{\rm -1}$)&(W Hz$^{\rm -1}$)&(W Hz$^{\rm -1}$)&percentage&($\mu$G)\\
\hline
\hline
NVSS &3.8$\times$10$^{\rm -3}$~$^*$&5.7$\times$10$^{\rm 38}$&~&~&2.5$\times$10$^{\rm 18}$&6.9$\times$10$^{\rm 18}$&1.2$\times$10$^{\rm 18}$&$\sim$8$\times$10$^{\rm 17}$&$\sim$70\%&$\sim$1.6\\
MIPS 70\um&3.0$\times$10$^{\rm -3}$&4.1$\times$10$^{\rm 38}$&1.3$\times$10$^{\rm 39}$&1.2$\times$10$^{\rm 40}$&~&~&~&~&~&~\\
common&3.0$\times$10$^{\rm -3}$&4.1$\times$10$^{\rm 38}$&1.4$\times$10$^{\rm 39}$&1.2$\times$10$^{\rm 40}$&1.2$\times$10$^{\rm 18}$&5.4$\times$10$^{\rm 18}$&9$\times$10$^{\rm 17}$&$\sim$7$\times$10$^{\rm 17}$&$\sim$40\%&$\sim$1.4\\
UGC 5456&1.9$\times$10$^{\rm -2}$&4.5$\times$10$^{\rm 39}$&2.4$\times$10$^{\rm 40}$&6.0$\times$10$^{\rm 40}$&1.1$\times$10$^{\rm 19}$&3.4$\times$10$^{\rm 19}$&8$\times$10$^{\rm 18}$&$\sim$6$\times$10$^{\rm 18}$&$\sim$50\%&$\sim$1.8\\
\hline
\hline
\end{tabular}
\end{center}
\begin{flushleft}
$^a$ The name of the sub-sample / galaxy; $^b$ the mean SFR for the sample galaxies as estimated from the measured FUV emission; $^c$ the estimated \ha\ luminosity; $^d$ the measured 70\um~luminosity; $^e$ the expected 70\um~luminosity for the listed \ha~flux; $^f$ the measured 1.4~GHz luminosity; $^g$ the expected 1.4~GHz luminosity for the given SFR (FUV based), estimated using the calibration in \citet{bel03} for L$>$L$_*$ galaxies; $^h$ the expected 1.4~GHz luminosity for the given SFR (FUV based), estimated using the calibration in \citet{bel03} for L$<$L$_*$ galaxies; $^i$ the estimated thermal emission at 1.4~GHz; $^j$ the estimated percentage flux of non-thermal origin at 1.4~GHz; $^k$ the estimated equipartition magnetic field.\\
$^*$ Mean value of the 46 galaxy `FUV' sub-sample.
\end{flushleft}
\end{table*}

From Table~\ref{tab:flux} it is evident that, within the measurement uncertainties the radio-FIR correlation holds for faint dwarf irregular galaxies.  However, it is unclear if the radio-FIR correlation continue to hold because the radio and FIR fluxes both  trace the SFR, or because of a `conspiracy' in which both of the tracers under-predict the star formation rate by about the same amount. To distinguish between these possibilities we now check how the FIR and radio fluxes separately compare to the fluxes that would be expected given their SFR and the radio and FIR SFR calibrations for large galaxies. The various relevant calculated quantities for the different sub-samples and UGC 5456, using the fluxes measured as described in the last section, are tabulated in Table~\ref{tab:cal}.
The SFRs listed in Column(2) were calculated using the calibrations given in \citet{ken98}, after accounting for foreground Galactic extinction using extinction values of \citet{sch98} and formulae from \citet{car89}. No correction for extinction due to dust within the galaxies were made, as such corrections are found to be small for the few brighter galaxies with detectable dust emission (dust fluxes and flux limits for a large number of FIGGS galaxies can be found in \citet{dal09}).
It is known that at low SFR the \ha\ emission is suppressed compared to what one would expect from the calibration between SFR and \ha\ flux at high SFR \citep{lee09,hun10,roy11}. The \ha\ luminosities listed in Column(3) are calibrated using the data for the FIGGS sample (i.e. at low SFRs) by \citet{roy11}.
The luminosities listed in Column(5) are estimated using the calibration of  \citet{cal10}, which is obtained from the fit to data for high metallicity (and more luminous) galaxies (their equation 18). As can be seen by comparing Columns(4) \& (5), our sample dwarfs have fainter 70$\mu$m luminosities than expected from this calibration. 
\citet{bel03} proposed separate empirical calibrations of SFR with 1.4~GHz radio luminosity, for galaxies with luminosity less or more than L$_*$. For low luminosity galaxies, it was assumed that there was a non linear relation between the 1.4~GHz flux and the SFR. The non-linearity is in the sense that at low SFRs, the 1.4~GHz flux gets disproportionately fainter.  \citet{bel03} proposed a particular functional form of this relation, chosen to ensure that the radio-FIR correlation would be satisfied. Column(7) gives the expected value using the calibration for luminous galaxies, while Column(8) uses this proposed calibration for faint galaxies. As can be seen the radio luminosities measured by us compare better with the calibration for faint (L$<$L$_*$) galaxies, i.e. are consistent with a non-linear relation between the SFR and the 1.4~GHz flux.
How the values listed in Columns(9) through (11) were arrived at, are discussed in detail below.

From the values listed in Table.~\ref{tab:cal} it is clear that our sample galaxies have fainter 70$\mu$ and $1.4$~GHz luminosities than expected from their SFR. In terms of the numerical modelling of \citet{lac10}, star-forming dwarf galaxies are neither radio nor FIR calorimeters. However both these quantities are suppressed by approximately the same amount, and hence the radio-FIR correlation continues to hold, i.e.  the `conspiracy' alluded to by \citet{bel03} appears to hold for the faintest star forming galaxies. It is interesting to note in this context, that the formula given in \citet{bel03} for converting the 1.4 GHz flux into a SFR appears to work reasonably well for our galaxies, even though the 1.4 GHz flux levels of our sample is about an order of magnitude lower than the that in his sample. 

As discussed above the 1.4~GHz flux of our sample galaxies appears suppressed. It would be interesting to check if this translates into a change in the ratio of thermal to non-thermal emission. For example if the magnetic fields in dwarf galaxies are weak and/or cosmic ray confinement is less as compared to large galaxies, one would expect the non-thermal emission to be suppressed. To estimate the ratio of non-thermal and thermal parts of the radio flux from the sources, we use \citet{cap86}'s method to estimate the thermal radio flux from the estimated \ha\ flux. \citet{cap86} give a relation to compute the expected 5~GHz flux from the observed \ha\ flux assuming case B recombination, electron density of 100 cm$^{\rm -2}$ and temperature of 10,000~K. We translate this to the expected 1.4~GHz flux assuming a spectral index of $\alpha$~= 0.1 (${\rm S~\sim~\nu^{-\alpha}}$) for the thermal emission. This estimated thermal contribution to the 1.4~GHz luminosity is listed in Col.(9) of Table~\ref{tab:cal}, while Col.(10) gives the corresponding non-thermal fraction of the total 1.4~GHz emission. For normal spirals, typically 90\% of the radio flux at 1.4 GHz is non-thermal in origin (\citet{con92}, \citet{nik97a}, \citet{bas12}). In contrast, for our sample galaxies it appears that a significant fraction $\gtrsim 50\%$ of the 1.4~GHz flux appears to be thermal in origin. We note that because of the uncertainties in the conversion from the FUV flux to the formation rate of massive stars \citep[see eg.][]{lee09,roy11} and hence the thermal radio flux the possibility that all of the observed flux comes thermal emission cannot be ruled out. Earlier studies have found thermal fractions higher than those in normal spirals, in local group starburst dwarfs like IC 10 \citep{chy03}, in the SMC \citep{loi87} and BCDs \citep{kle91}. Consistent with this, \citet{hee11} find that in IC10 the measured radio emission is much lower than what is expected from the SFR possibly due to escape of a substantial fraction of the CR electrons.
 
Next, from the non-thermal 1.4~GHz flux (i.e. the difference between the total flux and the estimated thermal flux), the equipartition magnetic field can be estimated using the formulae given in \citet{bec05}. The basic assumption behind this calculation is the equipartition of energy densities between the magnetic field and the cosmic rays. We make the following other assumptions about various parameters: the synchrotron emission spectral index in 0.7, the proton to electron number density is 100, and magnetic field is totally turbulent with all possible inclinations between the field and the sky plane at different locations in the galaxy. The synchrotron pathlength through the galaxy is taken to be the minor axis length. Dwarf galaxies have thick HI as well as stellar discs \citep{roy10,san10}, with mean axial ratio for the HI disk  $\sim 0.6$.  For the stacked sources, the mean Holmberg diameter as listed in Table~\ref{tab:samp} is hence multiplied by 0.6 to obtain the minimum synchrotron pathlength. The estimated equipartition magnetic fields are listed in Col.~(11) of Table~\ref{tab:cal}, and can be seen to be $\sim 2\mu$G.  This is about 5 times lower than the magnetic field strength of normal 
spirals \citep{beck05}.

\section{Summary}
\label{sec:sum}

We study the radio-FIR correlation for some of the faintest known star forming galaxies using radio and FIR flux estimates obtained by stacking the images of individual galaxies. We also use the observed SFR in these galaxies to estimate the expected amount of radio and FIR flux, assuming the SFR calibrations that hold for large galaxies. We estimate the expected thermal radio emission at 1.4~GHz using the observed SFR, as well as the equipartition magnetic field. Our findings are as follows: 

\begin{enumerate}
\item For their measured SFR both the 70\um~and 1.4 GHz flux for dwarf irregular galaxies is low compared to the calibration for large spirals.

\item The ratio of 70\um~to 1.4 GHz flux (q70) however agrees within errorbars with that seen for larger galaxies. The radio-FIR correlation thus appears to be the result of a `conspiracy' where both the 70\um~to 1.4 GHz decrease non-linearly with SFR compared to large spirals. Thus dwarf galaxies do not act as `calorimeters' for either CR electrons or UV photons. 

\item The inferred thermal fraction of the emission at 1.4~GHz for our sample galaxies is $\sim 50\%$, much larger than the 10\% typical for spirals. 

\item The estimated equipartition magnetic field is $\sim$ 2 $\mu$G, about 5 times smaller than the typical large-scale overall magnetic field in spirals.

\end{enumerate}

\bsp

\label{lastpage}


\begin{thebibliography}{}

\bibitem[\protect\citeauthoryear{Aniano et al.}{2011}]{ani11} Aniano G., Draine B. T., Gordon K. D., Sandstrom K., 2011, PASP, 123, 1218
\bibitem[\protect\citeauthoryear{Appleton et al.}{2004}]{app04} Appleton P. N. et al., 2004, ApJ Supp. Ser., 154, 147
\bibitem[\protect\citeauthoryear{Basu et al.}{2012}]{bas12} Basu Aritra, Mitra Dipanjan, Wadadekar Yogesh, Ishwara-Chandra C. H., 2012, MNRAS, 419, 1136
\bibitem[\protect\citeauthoryear{Beck}{2005}]{beck05} Beck R., 2005, in Cosmic Magnetic Fields, eds. R. Wielebinski, R. Beck, Heidelberg: Springer, 41
\bibitem[\protect\citeauthoryear{Beck \& Krause}{2005}]{bec05} Beck R. Krause M., 2005, Astron. Nachr., 326, No. 6, 414
\bibitem[\protect\citeauthoryear{Begum et al.}{2008}]{beg08} 
Begum A., Chengalur J.~N., Karachentsev I.~D., Sharina M.~E., Kaisin S.~S., 
2008, MNRAS, 386, 1667
\bibitem[\protect\citeauthoryear{Bell}{2003}]{bel03} Bell Eric F., 2003, ApJ, 586, 794
\bibitem[\protect\citeauthoryear{Bertone, Vogt \& Ensslin}{2006}]{ber06} Bertone S., Vogt C., Ensslin T., 2006, MNRAS, 370, 319
%\bibitem[\protect\citeauthoryear{Beswick et al.}{2008}]{bes08} Beswick R. J., Muxlow T. W. B., Thrall H., Richards A. M. S., 2008, MNRAS, 385, 1143
%\bibitem[\protect\citeauthoryear{Boyle et al.}{2007}]{boy07} Boyle B. J., Cornwell T. J., Middelberg E., Norris R. P., Appleton P. N., Smail Ian, 2007, MNRAS, 376, 1182
\bibitem[\protect\citeauthoryear{Calzetti et al.}{2010}]{cal10} Calzetti D. et al., 2010, ApJ, 714, 1256
\bibitem[\protect\citeauthoryear{Caplan \& Deharveng}{1986}]{cap86} Caplan J., Deharveng L., 1986, A\&A, 155, 297
\bibitem[\protect\citeauthoryear{Cardelli, Clayton \& Mathis}{1989}]{car89} Cardelli Jason A., Clayton Geoffrey C., Mathis John S., 1989, ApJ, 345, 245
\bibitem[\protect\citeauthoryear{Chyzy et al.}{2000}]{chy00} Chyzy K. T., Beck R., Kohle S., Klein U., Urbanik M., 2000, A\&A, 355, 128
\bibitem[\protect\citeauthoryear{Chyzy et al.}{2003}]{chy03} Chyzy K. T., Knapik J., Bomans D. J., Klein U., Beck R., Soida M., Urbanik M., 2003, A\&A, 405, 513
\bibitem[\protect\citeauthoryear{Chyzy et al.}{2011}]{chy11} Chyzy K. T., Wezgowiec M., Beck R., Bomans D. J., 2011, A\&A, 529, A94
\bibitem[\protect\citeauthoryear{Condon}{1992}]{con92} Condon J. J., 1992, ARA\&A, 30, 575
\bibitem[\protect\citeauthoryear{Condon et al.}{1998}]{con98} Condon J. J., Cotton W. D., Greisen E. W., Yin Q. F., Perley R. A., Taylor G. B., Broderick J. J., 1998, AJ, 115, 1693
\bibitem[\protect\citeauthoryear{Dale et al.}{2009}]{dal09} Dale D. A. et al., 2009, ApJ, 703, 517
%\bibitem[\protect\citeauthoryear{Garn \& Alexander}{2009}]{gar09} Garn Timothy, Alexander Paul, 2009, MNRAS, 394, 105
\bibitem[\protect\citeauthoryear{Gordon et al.}{2007}]{gor07} Gordon Karl D. et al., 2007, PASP, 119, 1019
\bibitem[\protect\citeauthoryear{Heesen et al.}{2011}]{hee11} Heesen V., Rau U., Rupen M. P., Brinks E., Hunter D. A., ApJ, 739, L23
\bibitem[\protect\citeauthoryear{Helou \& Bicay}{1993}]{hel93} Helou G., Bicay M. D., 1993, ApJ, 415, 93
\bibitem[\protect\citeauthoryear{Hunter, Elmegreen \& Ludka}{2010}]{hun10} Hunter Deidre A., Elmegreen Bruce G., \& Ludka Bonnie C., 2010, AJ, 139, 447
\bibitem[\protect\citeauthoryear{Kennicutt}{1998}]{ken98} Kennicutt Jr. Robert C., 1998a, ARA\&A, 36, 189
\bibitem[\protect\citeauthoryear{Kepley et al.}{2010}]{kep10} Kepley Amanda A., Mühle Stefanie, Everett John, Zweibel Ellen G., Wilcots Eric M., Klein Uli, 2010, ApJ, 712, 536
\bibitem[\protect\citeauthoryear{Klein, Weiland \& Brinks}{1991}]{kle91} Klein U., Weiland H., Brinks E., 1991, A\&A, 246, 323
\bibitem[\protect\citeauthoryear{Lacki, Thompson \& Quataert}{2010}]{lac10} Lacki Brian C., Thompson Todd A., Quataert Eliot, 2010, ApJ, 717, 1
\bibitem[\protect\citeauthoryear{Lee et el.}{2009}]{lee09} Lee Janice C. et al., 2009, ApJ, 706, 599
\bibitem[\protect\citeauthoryear{Li \& Draine}{2001}]{li01} Li Aigen, Draine B. T., 2001, ApJ, 554, 778
\bibitem[\protect\citeauthoryear{Loiseau et al.}{1987}]{loi87} Loiseau N., Klein U., Greybe A., Weilebinski R., Haynes R. F., 1987, A\&A, 178, 62
\bibitem[\protect\citeauthoryear{Mao et al.}{2008}]{mao08} Mao S. A., Gaensler B. M., Stanimirović S., Haverkorn M., McClure-Griffiths N. M., Staveley-Smith, L., Dickey J. M., 2008, ApJ, 688, 1029
\bibitem[\protect\citeauthoryear{Murgia et al.}{2005}]{mur05} Murgia M., Helfer T. T., Ekers R. Blitz L., Moscadelli L., Wong T., Paladino R., 2005, A\&A, 437, 389
%\bibitem[\protect\citeauthoryear{Niklas \& Beck}{1997}]{nik97} Niklas S., Beck R., 1997, A\&A, 320, 54
\bibitem[\protect\citeauthoryear{Niklas, Klein \& Wielebinski}{1997}]{nik97a} Niklas S., Klein U., Wielebinski R., 1997, A\&A, 322, 19
\bibitem[\protect\citeauthoryear{Padovani}{2011}]{pad11} Padovani Paolo, 2011, MNRAS, 411, 1547
\bibitem[\protect\citeauthoryear{Rieke et al.}{2004}]{rie04} Rieke G. H. et al., 2004, ApJ Supp. Ser., 154, 25
\bibitem[\protect\citeauthoryear{Roychowdhury et al.}{2010}]{roy10} Roychowdhury S., Chengalur J. N., Begum A., Karachentsev I. D. 2010, MNRAS, 404, L60
\bibitem[\protect\citeauthoryear{Roychowdhury et al.}{2011}]{roy11} Roychowdhury S., Chengalur J. N., Kaisin S. S., Begum A., Karachentsev I. D. 2011, MNRAS, 414, L55
\bibitem[\protect\citeauthoryear{Sanchez-Janssen, Mendez-Abreu \& Aguerri}{2010}]{san10} Sanchez-Janssen R., Mendez-Abreu J., Aguerri J. A. L., 2010, MNRAS, 406, L65
\bibitem[\protect\citeauthoryear{Schlegel, Finkbeiner \& Davis}{1998}]{sch98} Schlegel David J., Finkbeiner Douglas P., Davis Marc, 1998, ApJ, 500, 525
\bibitem[\protect\citeauthoryear{Thompson et al.}{2006}]{tho06} Thompson Todd A., Quataert Eliot, Waxman Eli, Murray Norman, Martin Crystal L., 2006, ApJ, 645, 186
\bibitem[\protect\citeauthoryear{Volk}{1989}]{vol89} Volk H. J., 1989, A\&A, 218, 67
\bibitem[\protect\citeauthoryear{Yun \& Reddy}{2001}]{yun01} Yun Min S., Reddy Naveen A., 2001, ApJ, 554, 803

\end{thebibliography}
\end{document}